\pdfoutput=1  
\documentclass[]{article}  
\usepackage{url,float}
\usepackage{graphicx}
\usepackage{amsmath}
\usepackage{amsfonts}
\usepackage{amssymb}
\usepackage{latexsym}

\newcommand{\hide}[1]{}

\newcommand{\ABox}{
\raisebox{3pt}{\framebox[6pt]{\rule{6pt}{0pt}}}
}
\newenvironment{proof}{{\bf Proof:}}{\hfill\ABox}

\newtheorem{theorem}{{\bf Theorem}}

\newtheorem{lemma}{Lemma}

\newcommand{\lemlab}[1]{\label{lemma:#1}}
\newcommand{\thmlab}[1]{\label{thm:#1}}

\newcommand{\figlab}[1]{\label{fig:#1}}
\newcommand{\seclab}[1]{\label{sec:#1}}

\newcommand{\lemref}[1]{\ref{lemma:#1}}
\newcommand{\thmref}[1]{\ref{thm:#1}}

\newcommand{\secref}[1]{\ref{sec:#1}}
\newcommand{\figref}[1]{\ref{fig:#1}}

{\makeatletter
 \gdef\xxxmark{%
   \expandafter\ifx\csname @mpargs\endcsname\relax 
     \expandafter\ifx\csname @captype\endcsname\relax 
       \marginpar{xxx}
     \else
       xxx 
     \fi
   \else
     xxx 
   \fi}
 \gdef\xxx{\@ifnextchar[\xxx@lab\xxx@nolab}
 \long\gdef\xxx@lab[#1]#2{{\bf [\xxxmark #2 ---{\sc #1}]}}
 \long\gdef\xxx@nolab#1{{\bf [\xxxmark #1]}}
 \gdef\turnoffxxx{\long\gdef\xxx@lab[##1]##2{}\long\gdef\xxx@nolab##1{}}%
}

\def\R{{\mathbb{R}}}
\def\D{{\Delta}}
\def\c{{\chi}}
\def\s{{\sigma}}

\title{%
A Note on Solid Coloring\\of Pure Simplicial Complexes
} 

\author{%
Joseph O'Rourke%
    \thanks{Department of Computer Science, Smith College, Northampton, MA
      01063, USA.
      \protect\url{orourke@cs.smith.edu}.}
}

\begin{document}
\maketitle

\begin{abstract}
We establish a simple generalization of a known result in the plane.
The simplices in any pure simplicial complex in $\R^d$ may be
colored with $d+1$ colors so that no two simplices that share a
$(d{-}1)$-facet
have the same color.
In $\R^2$ this says that any planar map all of whose faces
are triangles may be 3-colored,
and in $\R^3$ it says that tetrahedra in a collection
may be ``solid 4-colored'' so that no two glued face-to-face receive
the same color.
\end{abstract}

\section{Introduction}
\seclab{Introduction}
The famous 4-color theorem says that the regions of any planar map
may be colored with four colors such that no two regions that share a
positive-length border receive the same color.
A lesser-known special case is that if all the regions are triangles,
three colors suffice.
For the purposes of generalization, this can be phrased as building a
planar
object by gluing triangles edge-to-edge, and then 3-coloring the
triangles.
Because the coloring constraint in this formulation
only applies to triangles adjacent the
dual
graph---whose nodes are triangles and whose arcs join triangle nodes
that share a whole edge---slightly more general objects can be 3-colored:
\emph{pure} (or \emph{homogenous})\footnote{
  Pure/homogenous means that there are no dangling edges or isolated
  vertices, and in general, no pieces of dimension less than $d$ that
  are not part of a simplex of dimension $d$.  So the complex is a
  collection of $d$-simplices glued facet-to-facet.
}
\emph{simplicial complexes}
in $\R^2$, whose dual graph may have several components, with
independent colorings. 
See Figure~\figref{TriComplex}.
\begin{figure}[htbp]
\centering
\includegraphics[width=0.75\linewidth]{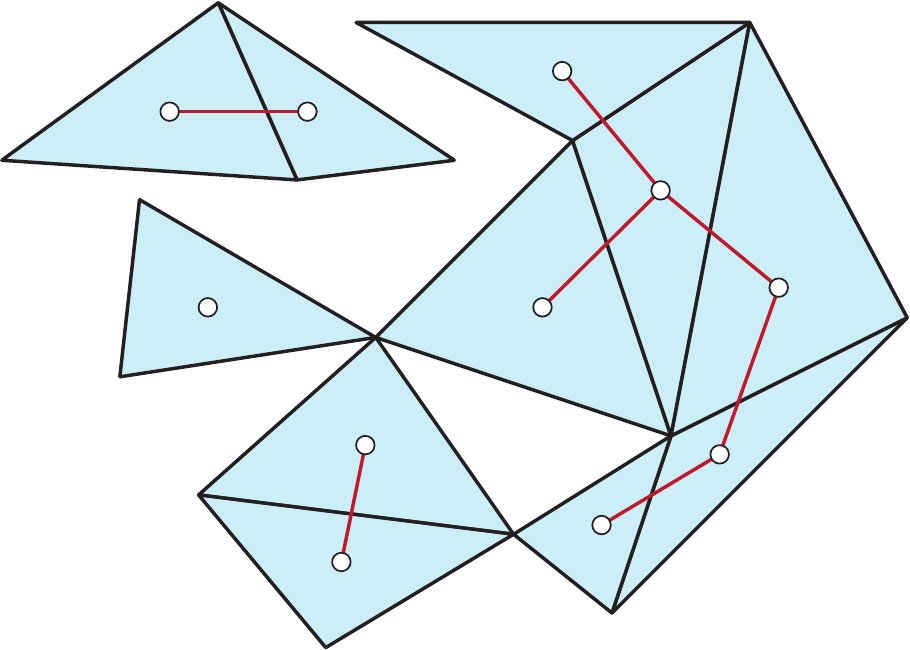}
\caption{A triangle complex and its dual graph $G$.}
\figlab{TriComplex}
\end{figure}

For simplicity, we will call such a complex a \emph{triangle complex},
its analog in $\R^3$ a \emph{tetrahedron complex},
and the generalization a \emph{$d$-simplex complex}.
We permit these complexes to contain an infinite number of simplices;
e.g., tilings of space by simplices are such complexes.
The main result of this note is:
\begin{theorem}
A $d$-simplex complex may be $(d{+}1)$-colored in the sense that each
simplex may be colored with one of $d+1$ colors
so that any pair of simplices that share
a $(d{-}1)$-facet receive different colors.
\thmlab{main}
\end{theorem}
One can think of the whole volume of each simplex being colored---so
``solid coloring'' of tetrahedra in $\R^3$.
Although I have not found this result in the literature,
it is likely known, as its proof is not difficult---essentially,
remove one simplex and induct.
Consequently, this
note should be considered expository,
and I will describe proofs in more detail than in a research announcement.
Perhaps more interesting than the result itself 
are the many related questions in Section~\secref{Beyond}.

\section{Triangle Complexes}
\seclab{TriangleComplexes}
Let $G$ be the dual graph of a triangle complex, and
let $\D(G) = \D$ be the maximum degree of nodes of $G$.
For triangle complexes, $\D=3$.
Let $\c(G) = \c$ be the chromatic number of $G$.
An early result of Brooks~\cite{b-ocnn-41}
says that
$\c \le \D + 1$ for any graph $G$.
For duals of triangle complexes, this theorem only
yields $\c=4$, the 4-color theorem for triangle complexes.
We now proceed to establish $\c=3$ in three stages:
\begin{enumerate}
\item We first prove it for finite triangle complexes.
\item We then apply a powerful result of deBruijn and  Erd\H{o}s to
  extend
the result to infinite complexes.
\item We formulate a second proof for infinite complexes that does not
invoke deBruijn-Erd\H{o}s.
\end{enumerate}
The primary reason for offering two proofs is that related questions
raised
in Section~\secref{Beyond} may benefit from more than one proof
approach.

\subsection{Finite Triangle Complexes}
Let $S$ be a triangle complex containing a finite number of triangles,
and $G$ its dual graph.
Let $C(S)=C$ be the convex hull of $S$, i.e., the boundary of the
smallest
convex polygon enclosing $S$.
The proof is by induction on the number of triangles, with
the base case of one triangle trivial.
\begin{description}
\item[Case 1.] There is a triangle $t$ with at least one edge $e$ on
  $C$.
Then $e$ is \emph{exposed}
(i.e., not glued to another
triangle of the complex),
and $t$ has at most degree 2 in $G$.
Remove $t$ to produce complex $S'$, 3-color $S'$ by induction,
put back $t$, and color it with a color distinct from the colors of
its
at most 2 neighbors in $G$.
\item[Case 2.] No triangle has an edge on $C$.  Let $v$ be any vertex
  of $C$, and let $t$ be the most counterclockwise (ccw) triangle incident
  to $v$.
See Figure~\figref{TriangleHull}.
\begin{figure}[htbp]
\centering
\includegraphics[width=0.75\linewidth]{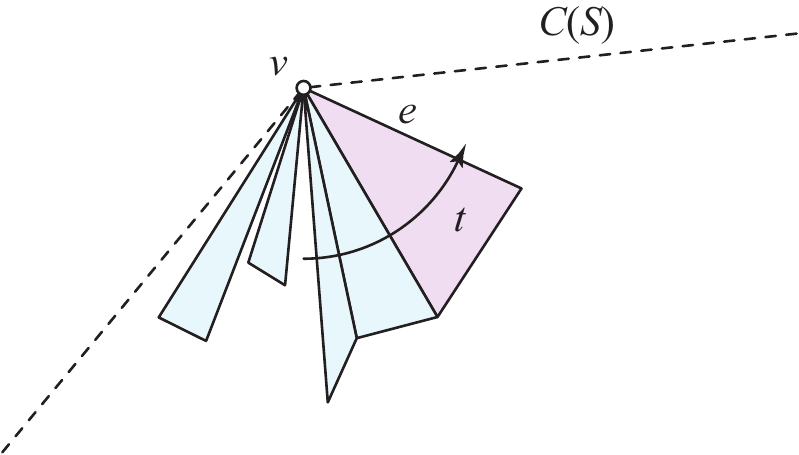}
\caption{Triangle $t$ has an exposed edge $e$.}
\figlab{TriangleHull}
\end{figure}
Then the ccw edge $e$ of $t$ incident to $v$ is exposed.
Then---just as in the previous case---remove
$t$,
3-color by induction, put $t$ back colored with a color not used by
its
at most two neighbors.
\end{description}
This simple induction argument establishes $\c=3$ for finite triangle complexes.

\subsection{deBruijn-Erd\H{o}s}

The result of deBruijn and  Erd\H{o}s is this~\cite{ed-cpigp-51}:
\begin{theorem}
If a graph $G$ has the property that any finite subgraph is
$k$-colorable,
then $G$ is $k$-colorable itself.
\thmlab{deB-E}
\end{theorem}

This immediately extends the result just proved to infinite triangle
complexes.
Note that the induction proof presented fails for infinite complexes, 
because it is possible that every
triangle
has degree 3 in $G$ for infinite complexes,
for example, in a triangular tiling.

\subsection{Proof based on $K_r$}
The alternative proof in some sense ``explains'' why a triangle complex
is
3-colorable: because it does not contain $K_4$ as a subgraph.
Of course we could obtain this indirectly by using the above proof and
conclude that $K_4$ could not be a subgraph (because it needs 4
colors),
but establishing it directly gives additional insight.

We rely here on this result,
obtained independently by several researchers
(Borodin and Kostochka, Catlin, and Lawrence, as reported in~\cite{s-nubcn-02}):
\begin{lemma}
If $G$ does not contain any $K_r$ as a subgraph,
$4 \le r \le \D + 1$,
then 
$$\c \le \frac{r-1}{r} (\D + 2) \;.$$
\lemlab{Lawrence}
\end{lemma}
We will now show that $K_4$ is not a subgraph of $G$ for triangle
complexes,
which, because $r=4$ and $\D=3$, then implies
$$ \c \le \frac{3}{4} (3 + 2) = 3\frac{3}{4} \;,$$
and so (because $\c$ is an integer), $\c \le 3$.

\begin{lemma}
$K_4 \not\subseteq G$.
\lemlab{notK4}
\end{lemma}
\begin{proof}
\emph{Sketch.}
We only sketch the argument, because in the Appendix we prove
more formally the extension to $\R^d$, including $d=2$.

\begin{figure}[htbp]
\centering
\includegraphics[width=0.5\linewidth]{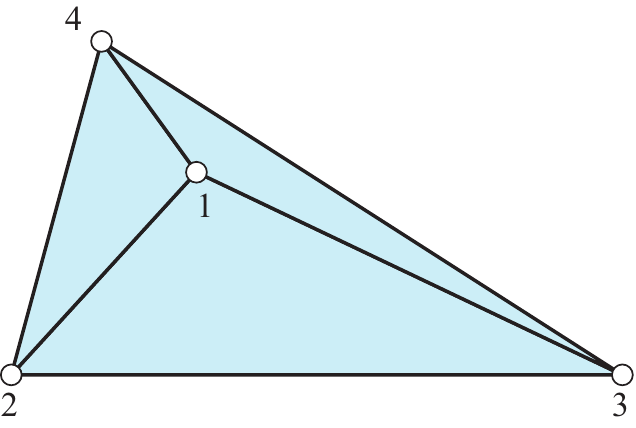}
\caption{Triangles forming $K_3$.}
\figlab{TriK3}
\end{figure}
If $K_4$ is a subset of $G$, then $K_3$ must be as well.
The only configuration of triangles that realizes $K_3$ is that shown
in Figure~\figref{TriK3}: the three triangles share and surround a vertex (labeled
1 in the figure).  Now consider attempting to
extend this to $K_4$ by gluing another triangle to the only uncovered
edge of $\triangle \{1,2,3\}$, edge $e=\{2,3\}$.  Its apex, call it
$v_5$,
must lie below $e$, but because $v_4$ lies above $e$,
the new triangle $\triangle \{2,3,5\}$ cannot share the edges
$\{2,4\}$ and $\{3,4\}$, which it must to be adjacent to the other two
triangles.  Therefore, $K_4$ cannot occur in $G$, and we have established the claim.
\end{proof}

And as we argued above, Lemmas~\lemref{Lawrence} and~\lemref{notK4}
together
imply that $\c( G ) \le 3$: triangle complexes are 3-colorable.

\section{Tetrahedron Complexes}
Again we follow the same procedure as above, although we will defer
consideration of $K_5$ to general $d$-simplex complexes to the
Appendix, Section~\secref{Kd2}.
Now $S$ is a finite tetrahedron complex, $G$ its dual graph, and $C$
the convex hull of $S$, the boundary of a convex polyhedron.
Again the proof is by induction.  Although we could repeat the
structure
of the proof for triangle complexes, we opt for an argument that more
easily
generalizes to $d$ dimensions.

Let $v$ be a vertex of the hull $C=C(S)$, and let $S_v$ be the subset of
$S$ of tetrahedra incident to $v$.
Let $C_1=C(S_v)$ be the convex hull of $S_v$.
If there is a tetrahedron $t \in S_v$ with at least one face $f$ lying
on $C_1$, then $t$ has at most 3 neighbors in $S$.  Remove $t$,
4-color the smaller complex $S'$, put $t$ back, and color it with a
color not used for its at most 3 neighbors.
Note that it could well be that the face $f$ lies on $C(S)$ because
$C_1$ and $C$ coincide at $f$.  But having $f$ on $C$ is not the
crucial fact; if it is on $C_1$, it is exposed, and induction then
applies.

If no tetrahedron in $S_v$ has a face on $C_1$, then there must be a
tetrahedron
$t$ that has an edge $e$ on $C_1$ 
(in fact, there must be at least three such tetrahedra).
See Figure~\figref{TetraHull}.
\begin{figure}[htbp]
\centering
\includegraphics[width=0.5\linewidth]{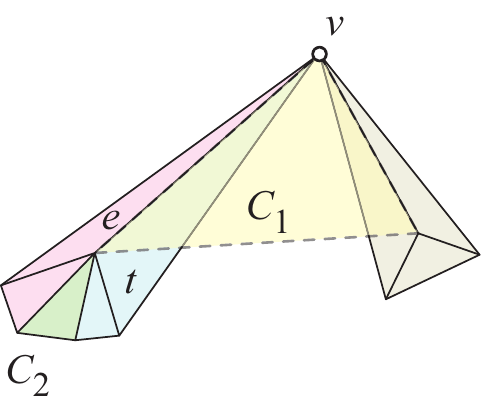}
\caption{No tetrahedron has a face on $C_1$.}
\figlab{TetraHull}
\end{figure}

Let $S_e$ be the subset of those tetrahedra in $S_v$ that share $e$.
Let $C_2=C(S_e)$ be the convex hull of these tetrahedra.
It must be that at least one tetrahedron has a face on $C_2$.
The tetrahedra sharing $e$ are angularly sorted about $e$, and we can
select
the most ccw one (which might be the same as the most cw one if
$|S_e|=1$).
So we have identified a tetrahedron with an exposed face, and
induction applies and establishes the result:
finite tetrahedon complexes have $\c=4$.
Infinite tetrahedon complexes follow from 
Theorem~\thmref{deB-E}.
And we could now work backward to conclude that $G$ cannot contain
$K_5$ as a subgraph.

\section{$d$-Simplex Complexes}
We repeat the outline just employed.
The only difficult part is showing that in a finite $d$-simplex
complex $S$, there must be a simplex with an exposed facet.\footnote{
   We use \emph{facet} for a $(d{-}1)$-dimensional face, and
   \emph{face} for any smaller dimensional face.
}
Then induction goes through just as before.

Say that a convex hull $C$ of points in $d$ dimensions is \emph{full-dimensional}
if $C$ is not contained in a $(d{-}i)$-dimensional flat (hyperplane)
for any $i > 0$.

Let $v$ be a vertex of the hull $C=C(S)$, and let $S_v$ be the subset of
$S$ of simplices incident to $v$.
Let $C_1=C(S_v)$ be the convex hull of $S_v$;
this is a $d$-polytope that contains $S_v$.
If there is a simplex $\s \in S_v$ with at least one
$(d{-}1)$-dimensional facet $f$ 
contained in $C_1$, then $\s$ has at most $d$ neighbors in $S$,
and induction establishes that $S$ may be $(d{+}1)$-colored.

So suppose that no simplex in $S_v$ has a $(d{-}1)$-dimensional facet on $C_1$.
Let $|S_v|=n$.
We must have $n>1$, because otherwise $C_1$ would bound a single
simplex,
and all of its facets would be on $C_1$ and so exposed.
We know $C_1$ is full dimensional because it contains $d$-simplices.
Let $\s_1 \in S_v$ be a simplex that has a $k$-dimensional face $f_1$ in
$C_1$, such that
$k < d-1$ is maximal among all simplices with faces in $C_1$.
We claim that there must be another simplex $\s' \in S_v$ that also has a face $f'$ in
$C_1$, where $f' \neq f_1$.
For suppose otherwise, that is, suppose that all simplices in $S_v$
share $f_1$.  Then, because $C_1$ is full-dimensional,
one of these simplices $\s''$ must have a vertex $u$ not part of $f_1$
on $C_1$
(otherwise all simplices lie in the flat containing $f_1$).
But then $\s''$ has a face (the hull of $u$ and $f_1$)
on $C_1$ of dimension larger than $k$,
contradicting the choice of $\s_1$.

So $\s'$ has a face on $C_1$, and $\s'$ does not share $f_1$.
Let $S_{f_1}$ be all the simplices in $S_v$ that share
$f_1$, and
let $C_2$ be the convex hull of $S_{f_1}$.
Because we know that $\s' \not\in S_{f_1}$,
$|S_{f_1}| < n$.

Now the argument is repeated: $C_2$ is full-dimensional because it
includes at least one $d$-simplex $\s_1$.
If some simplex in $S_{f_1}$ has a $(d{-}1)$-dimensional facet on
$C_2$,
we have identified an exposed face.
Otherwise, we select some simplex $\s_2$ with a face $f_2$ on $C_2$,
and separate out into $S_{f_2}$ all the simplices sharing $f_2$.
$S_{f_2}$ must have at least one fewer simplex than does $S_{f_1}$,
following the same reasoning.

Continuing in this manner, we identify smaller and smaller subsets of
$S$:
$$
|S| \ge |S_v| > |S_{f_1}| > |S_{f_2}| > \cdots
$$
via repeated convex hulls $C_1, C_2, \ldots$,
and eventually either identify a simplex with a $(d{-}1)$-dimensional
facet on
the corresponding hull $C_i$, or reach a set of one simplex, which
has all of its facets exposed.
So there is always a simplex with an exposed facet:

\begin{lemma}
Any finite $d$-simplex complex contains a simplex with an 
exposed  $(d{-}1)$-dimensional
facet.
\lemlab{exposed}
\end{lemma}
\noindent
Given the nearly obvious nature of
this lemma, it seems likely there is a less labored proof
that identifies an exposed simplex more directly.

This lemma then proves Theorem~\thmref{main} for finite complexes,
and deBruijn-Erd\H{o}s establishes it for infinite complexes.
Again we may now conclude that $K_{d+2}$ cannot be a subgraph
of $G^{(d)}$,
where
we use the notation $G^{(d)}$ for the dual graph of a $d$-simplex
complex.
A geometric proof of this non-subgraph result is offered in the
Appendix.
With that, we obtain an alternative proof of  Theorem~\thmref{main},
which we restate in slightly different notation:
\begin{theorem}
The dual graph $G^{(d)}$ of a $d$-simplex complex in $\R^d$
has chromatic number $\c \le d+1$.
\end{theorem}
\begin{proof}
Lemma~\lemref{notKd2} tells us that $K_r$ is not a subgraph of
$G=G^{(d)}$,
with $r=d+2$.
We have that $\D = d+1$ because each $d$-simplex has $d+1$ facets.
Therefore we have 
$$4 \le r=d+2 \le \D + 1 = d+2 \;$$
for $d \ge 2$.
Therefore
Lemma~\lemref{Lawrence} applies, and yields
$$\c \le \frac{d+1}{d+2} (d+3) \;.$$
Now we can see that 
$$\frac{d+1}{d+2} (d+3) < d+2$$
by expanding $(d+1)(d+3)$ and $(d+1)^2$:
$$d^2 + 4d + 3 < d^2 + 4d + 4 \;.$$
Thus $\c$ is strictly less than $d+2$, which, because $\c$ is an
integer,
implies $\c \le d+1$.
\end{proof}

\section{Beyond Simplices}
\seclab{Beyond}
One can ask for analogs of Theorem~\thmref{main} for complexes composed
of shapes beyond simplices.
In the plane, a natural generalization is a complex built from
convex quadrilaterals glued edge-to-edge.
These complexes sometimes need four colors, as
the example in 
Figure~\figref{ConvexQuads} shows.
\begin{figure}[htbp]
\centering
\includegraphics[width=0.5\linewidth]{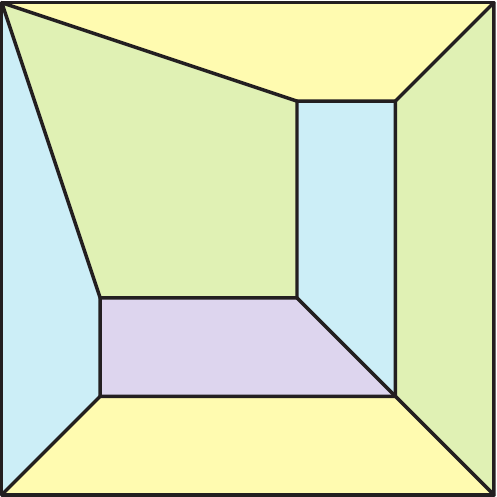}
\figlab{ConvexQuads}
\caption{A convex quadrilateral complex that needs four colors
\protect\cite[Fig.3a]{sw-rpt3c-00}.
}
\end{figure}
One does not need the 4-color theorem for this restricted class,
even without the convexity assumption:
there must exist a quadrilateral in a quadrilateral complex
with an exposed edge, and
4-coloring follows by induction.
Complexes built from pentagons 
can be proved
4-colorable by modifying the Kempe-chain argument;\footnote{
   I owe this observation to Sergey Norin,
   \url{http://mathoverflow.net/questions/49743/4-coloring-maps-of-pentagons}.
} 
so again
the full 4-color theorem is not needed here.


Sibley and Wagon proved in~\cite{sw-rpt3c-00}
the beautiful result: if the convex quadrilaterals are all parallelograms,
then three colors suffice
(essentially because there must be a parallelogram with two exposed edges).
In particular, Penrose rhomb tilings (their original interest) are 3-colorable.
Even more restrictive is requiring that the parallelograms be
rectangles.  Here with a student I proved  in~\cite{go-cobb-03} that such
\emph{rectangular brick} complexes of genus 0 are 2-colorable.
It is easily seen that complexes of genus 1 or greater might need
three colors (surround a hole with an odd cycle).

We also explored generalizations to $\R^3$ in~\cite{go-cobb-03}.
Somewhat surprisingly, genus-0 complexes built from \emph{orthogonal
bricks} (rectangular boxes in 3D) are again 2-colorable.
We also established that genus-1 orthogonal brick complexes are
3-colorable,
and conjectured that the same result holds for arbitrary genus.
I am aware of no substantive results on complexes built from
parallelopipeds (aside from the observation in~\cite{go-cobb-03} that
four colors are sometimes necessary), 
a natural generalization of the Sibley-Wagon result.\footnote{
  Our attempted proof in~\cite{go-cobb-03} for zonohedra is flawed.
}
One could also generalize convex quadrilaterals to convex hexahedra
(distorted cubes).
All of these generalizations seem unexplored.

\section{Appendix: $K_{d+2} \not\subseteq G^{(d)}$}
\seclab{Kd2}
Here we establish that $K_{d+2} \not\subseteq G^{(d)}$ without appeal
to deBruijn-Erd\H{o}s.
We partition the argument into four lemmas, the first three of which
show that there is essentially only one configuration that
achieves $K_{d+1}$, the analog of the configuration in
Figure~\figref{TriK3}.
The fourth lemma then shows that $K_{d+2}$ cannot be achieved.

Let $\s_1$, $\s_2$, and $\s_3$ be $d$-simplices.
Suppose $\s_1$ and $\s_2$ share a $(d{-}1)$-facet.
We will represent each simplex by the set of its vertex labels,
with distinct labels representing distinct points in $\R^d$.
When specifically referring to the point in space corresponding
to label $i$, we'll use $v_i$.
Let $\s_1= \{ 1,2,\ldots,d,(d{+}1) \}$
$\s_2= \{ 1,2,\ldots,d,(d{+}2) \}$,
with $\s_1 \cap \s_2 = f_{12}=\{ 1,2,\ldots,d \}$ their shared $(d{-}1)$-facet.
Under these circumstances, the following lemma holds:

\begin{lemma}
If $\s_3$ shares a $(d{-}1)$-facet with $\s_1$ and 
a $(d{-}1)$-facet
with $\s_2$ (and so the three simplices form $K_3$ in the dual),
then the $d+1$ vertices of $\s_3$ are among the $d+2$ vertices
of $\s_1 \cup \s_2 = \{ 1,2,\ldots,d,(d{+}1),(d{+}2) \}$:
$\s_3$ cannot include a vertex that is not a vertex of 
either $\s_1$ or $\s_2$.
\lemlab{K3.d2}
\end{lemma}
\begin{proof}
Suppose to the contrary that $\s_3$ includes a new vertex labeled $(d{+}3)$.
For $\s_3$ to share a $(d{-}1)$-facet with $\s_1$,
it needs to match $d$ of the $d+1$ vertices of $\s_1$.
But it cannot match the facet $f_{12} =\{ 1,2,\ldots,d \}$ because
that is already covered by $\s_2$.
Without loss of generality, let us assume that $\s_3$ includes
vertex $(d{+}1)$ but excludes vertex $k$ with $1 \le k \le d$.
So the $d+1$ vertices of $\s_3$ are
$$\s_3 = \{ (d{+}1), 1,2,\ldots,(k{-}1),(k{+}1),\ldots,d,(d{+}3)\} \;.$$
Now comparison to $\s_2$,
$$\s_2= \{ 1,2,\ldots,d,(d{+}2) \} $$
shows that it is not possible for $\s_3$ to match $d$ of the $d+1$
vertices of $\s_2$ (as it must to share a $(d{-}1)$-facet):
the two only share $d-1$ labels: 
$$\s_2 \cap \s_3 = \{1,2,\ldots,(k{-}1),(k{+}1),\ldots,d \}\;.$$
This contradiction establishes the claim.
\end{proof}

\begin{lemma}
Suppose $d+1$ $d$-simplices are glued together so that their dual
graph is $K_{d+1}$.
Then all the simplices together include only $d+2$ vertices.
\lemlab{Kd1.d2}
\end{lemma}
\begin{proof}
Let $\s_1,\ldots,\s_{d+1}$ be the simplices. 
By Lemma~\lemref{K3.d2}, $\s_1,\s_2,\s_3$ together include only $d+2$
vertices,
the $d+2$ vertices of $\s_1 \cup \s_2$.
But then repeating the argument for $\s_i$ for each $i=4,5,\ldots,d+1$
yields the same conclusion.
\end{proof}

\begin{figure}[htbp]
\centering
\includegraphics[width=0.75\linewidth]{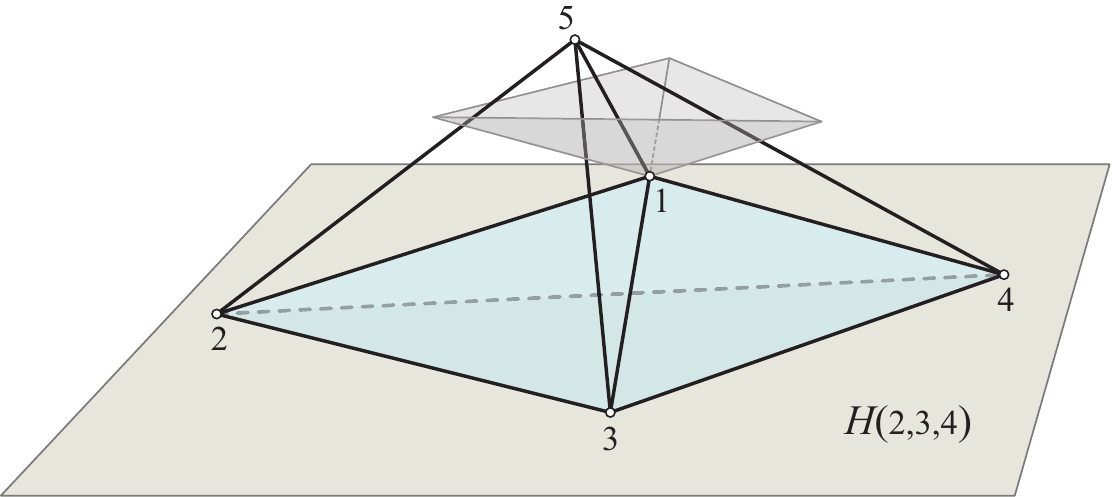}
\caption{Four tetrahedra whose dual forms $K_4$.}
\figlab{TetraLemma}
\end{figure}

We continue to study the $K_{d+1}$ configuration in the above
lemma.  Let us specialize to $d=3$ to make the situation clear.
We have four tetrahedra glued together to form $K_4$,
and Lemma~\lemref{Kd1.d2} says they have altogether 5 vertices.
Because $\binom{5}{4}=5$, only one of the possible combinations
of the labels $\{1,2,3,4,5\}$ is missing among the four tetrahedra.
Without loss of generality, we can say that $\{2,3,4,5\}$ is missing,
and that our four tetrahedra have these labels:

$$\{  1, 2, 3, 4  \}$$
$$\{  1, 2, 3, 5  \}$$
$$\{  1, 2, 4, 5  \}$$
$$\{  1, 3, 4, 5  \}$$

Our next claim is that $v_5$ lies to the same
side of the plane determined by the face $\{ 2,3,4 \}$ as does $v_1$.
Refer to Figure~\figref{TetraLemma}.

Let $H(i,j,k)$ be the plane containing the vertices with labels
$i$, $j$, and $k$.
Let $H^+(i,j,k;m)$ be the open halfspace bound by $H(i,j,k)$
and exterior to the tetrahedron $\{i,j,k,m\}$,
and $H^-(i,j,k;m)$ the analogous open halfspace including tetrahedron $\{i,j,k,m\}$.
The claim is that $v_5 \in H^-(2,3,4;1)$. 
The other three tetrahedra can each be viewed as the hull of $v_5$ and
one of the three faces of the $\{1, 2, 3, 4 \}$ tetrahedron above
the base: $\{1,2,3\}$, $\{1,2,4\}$, and $\{1,3,4\}$.
Because a tetrahedron can only be formed by a point above
each of these faces, we have that
$$v_5 \in H^+(1,2,3;4)$$
$$v_5 \in H^+(1,2,4;3)$$
$$v_5 \in H^+(1,3,4;2)$$
So $v_5$ must lie in the intersection of these three halfspaces,
which is a cone apexed at $v_1$ that is strictly above
the base plane $H(2,3,4)$.  See again Figure~\figref{TetraLemma}.
And therefore $v_5 \in H^-(2,3,4)$,
as claimed.

We now repeat this argument for $d$-simplices, where
the logic is identical but is perhaps obscured by the notation.

The configuration of $d+1$ $d$-simplices forming $K_{d+1}$ in
Lemma~\lemref{Kd1.d2} uses only $d+2$ vertices.
Because $\binom{d+2}{d+1}=d+2$, only one of the combinations
of $d+1$ labels is missing, which we take to be 
$\{2,3,\ldots,(d{+}2)\}$ without loss of generality.
So the labels of the $d+1$ simplices are:
$$\{ 1,2,\ldots,d,(d{+}1) \}$$
$$\{ 1,2,\ldots,d,(d{+}2) \}$$
$$\{ 1,2,\ldots,(d{+}1),(d{+}2) \}$$
$$\cdots$$
$$\{ 1,3,\ldots,d,(d{+}1),(d{+}2) \}$$

\begin{lemma}
In the configuration of $d+1$ simplices forming $K_{d+1}$ labeled as
just detailed above, $v_{d+2}$ lies in $H^- = H^-(2,3,\ldots,(d{+}1);1)$,
the same halfspace in which $v_1$ lies.
\lemlab{Kd1}
\end{lemma}
\begin{proof}
$H(2,3,\ldots,(d{+}1) )$ is the flat containing the ``base'' of the first simplex in the
list above, $\s_1 = \{ 1,2,\ldots,d,(d{+}1) \}$. 
The remaining $d$ simplicies in the list share the facets of $\s_1$
incident to $v_1$, each including $v_{d+2}$.
Thus $v_{d+2}$ is above each of those facets, i.e., it lies in the
corresponding $H^+$ halfspaces:
$$v_{d+2} \in H^+( 1,2,\ldots,d; (d{+}1) )$$
$$\cdots$$
$$v_{d+2} \in H^+(1,3,\ldots,d,(d{+}1); 2  )$$
And therefore $v_{d+2}$ lies in the intersection of all these halfspaces,
which is a cone apexed at $v_1$ and lying strictly above $H(2,3,\ldots,(d{+}1))$.
Therefore $v_{d+2}$ is in $H^-$.
\end{proof}

Completing the argument is now straightforward.

\begin{lemma}
$K_{d+2} \not\subseteq G^{(d)}$
\lemlab{notKd2}
\end{lemma}
\begin{proof}
Assume to the contrary that $K_{d+2}$ is a subgraph of $G^{(d)}$.
Then $K_{d+1}$ must be also.
Using the notation of Lemma~\lemref{Kd1},
that lemma establishes that in a configuration that
realizes $K_{d+1}$, vertex $v_{d+2}$ lies in 
$H^- = H^-(2,3,\ldots,(d{+}1); 1 )$.
Because $\{ 2,3,\ldots,(d{+}1) \}$ is the only facet of
the simplex  $\s_1 = \{ 1,2,\ldots,d,(d{+}1) \}$ not yet covered by
another simplex, the last simplex $\s_{d+2}$ must have labels
$\{ 2,\ldots,d,(d{+}1),(d{+}2) \}$.
And therefore $v_{d+2} \in H^+(2,3,\ldots,(d{+}1); 1)$.
But this is a contradiction, as it is saying that $v_{d+2}$ must lie
strictly to both sides of $H(2,3,\ldots,(d{+}1))$.
\end{proof}


\bibliographystyle{alpha}
\bibliography{/Users/orourke/bib/geom/geom}
\end{document}